\documentclass[aps,prd,onecolumn,nofootinbib,showkeys,amssymb,11pt]{revtex4}
\usepackage[a4paper,left=20mm,right=20mm,top=25mm,bottom=25mm]{geometry}
\usepackage{amsmath}
\usepackage{amsfonts}
\usepackage{amssymb}
\usepackage[utf8]{inputenc}
\usepackage[T1]{fontenc}
\usepackage[colorlinks=true]{hyperref}
\usepackage[dvipsnames]{xcolor}
\usepackage{graphicx}
\usepackage[normalem]{ulem}
\usepackage{float}
\usepackage{amssymb}
            
\begin{document}

\title{On Nash theory of gravity with matter contents}

\author{Phongpichit Channuie}
\affiliation{ College of Graduate Studies, Walailak University, Thasala, Nakhon Si Thammarat, 80160, Thailand}
\affiliation{School of Science, Walailak University, Nakhon Si Thammarat, 80160, Thailand}
\affiliation{Research Group in Applied, Computational and Theoretical Science (ACTS), \\Walailak University, Nakhon Si Thammarat, 80160, Thailand}

\author{Davood Momeni} 
 \affiliation{{Department of Physics, College of Science, Sultan Qaboos University,\\P.O. Box 36, Al-Khodh 123,  Muscat, Sultanate of Oman}}  

\author{
Mudhahir Al Ajmi
}

 \affiliation{{Department of Physics, College of Science, Sultan Qaboos University,\\P.O. Box 36, Al-Khodh 123,  Muscat, Sultanate of Oman}} 
\begin{abstract}

One of the alternative theories to Einstein’s general theory, a  divergence-free theory was proposed by J. Nash with Lagrangian density given by $2R_{\mu\nu}R^{\mu\nu}-R^2$. Although it was proved that the Nash theory doesn't have classical Einstein limits, it has been proven to be formally divergent free and considered to be of interest in constructing theories of quantum gravity. The original Nash gravity without matter contents can't explain the current acceleration expansion of the Universe. A possible extension of theory is by adding some matter contents to the model. In this work, we generalize Nash theory of gravity by adding the matter fields. In order to examine the effects of this generalization, we first derive the equations of motion in the flat FLRW spacetime and examine the behaviors of the solutions by invoking specific forms of the Hubble parameter. We also classify the physical behaviors of the solutions by employing the stability analysis and check the consistency of the model by considering particular cosmological parameters.
 
\end{abstract}

\date{\today}

\maketitle

\section{Introduction}
Several cosmological observations convince us that the observable
universe is undergoing a phase of accelerated expansion. The idea of a short period of extremely rapid expansion when the universe was very young is much attractive. However, the mechanism underlying this behavior is still a mystery. Cosmic inflationary scenario deserves one of the well-accepted mechanisms in order to describe the expansion Ref.\cite{Starobinsky:1980te}. Nowadays, inflationary cosmology is an inevitable ingredient when describing the very early evolution period of the universe. The reason is that it not only solve most of the puzzles that plague the standard Big Bang theory, but its prediction is also simultaneously consistent with the observations. Moreover, the discovery of the accelerating expansion of the Universe through observations of
distant supernovae mimics the current expansion of the universe \cite{SN,SN1}. In addition, various cosmological observations concede such an expansion. These include cosmic microwave background (CMB) radiation~\cite{Ade:2015xua,Ade:2015xua1,Ade:2015xua2,Ade:2015xua3,Ade:2015xua4,Ade:2015xua5,Ade:2015xua6}, large scale structure~\cite{LSS,LSS1}, baryon acoustic oscillations (BAO)~\cite{Eisenstein:2005su} as well as weak lensing~\cite{Jain:2003tba}. 

Regarding the late-time cosmic expansion, there are several possible explanations. Intuitively, the first one is the introduction of the dark energy component in the universe. However, the dark energy sector of the universe still remains an open question. In addition,
the second scenario is to interpret this unknown phenomenon by utilizing a purely geometrical picture. The latter is well known as the modified theory of gravity. Interestingly, modified theories of gravity have received more attraction lately due to numerous motivations ranging from high-energy physics, cosmology and astrophysics \cite{od1,od3}.

The $f(R)$ gravity constitutes one of the simplest versions of such modification. Several version of $f(R)$ gravity have been proposed and investigated so far, see for example Refs.\cite{Buc,Buc1,Buc2}. Here the Lagrangian density is an arbitrary function of the scalar curvature, $R$. For more details, see comprehensive reviewed articles on $f(R)$ theories \cite{fr}.\footnote{ Note that the
first model that can describe cosmic inflation was proposed by Starobinsky \cite{Starobinsky:1980te}.} Notice that the modified $f(R)$ gravity gives
good explanation for the cosmic acceleration without invoking the dark energy component implied from the cosmological data. Another interesting extension of the GR proposed in the form of $f(R,R_{\mu\nu}R^{\mu\nu})$ \cite{Bogdanos:2009tn}. 

 Among numerous different functional forms for $f(R,R_{\mu\nu}R^{\mu\nu})$ a divergence free form as $f(R,R_{\mu\nu}R^{\mu\nu})=2R_{\mu\nu}R^{\mu\nu}-R^2$ eventually coincides with a toy model for gravity which was introduced by  Nash. This specific form of model is divergence free and it could be considered as  a modified theory of gravity for empty space by employing higher-derivatives instead of the usual Einstein-Hilbert action \cite{Nash}. There are few number of papers about this interesting theory which was rarely investigated in literature \cite{Aadne:2017oba}-\cite{Channuie:2018now}.

In this work, we anticipate to generalize this later  theory by adding the matter fields in the original action. We specify a proper form of the field equations on more general footings for space with matter contents. In Sec.(\ref{2}), we generalize  theory of gravity by including the matter field in the original action. In addition, we derive the equations of motion in the flat FLRW spacetime and examine the behaviors of the solutions by invoking specific forms of the Hubble parameter. We also classify the physical behaviors of the solutions by employing the stability analysis. In Sec.(\ref{om}), we check the consistency of the model by considering cosmological parameters, e.g., the Hubble parameter $H$,  deceleration 
parameter $q$, and \emph{Om}(z) parameter. Finally, we conclude our findings in the last section.

\section{Divergence free Nash theory gravity with matter contents}
\label{2}
A significant feature of divergence free Nash theory of gravity is that the scalar curvature term $R$ in four-dimensional Riemanninan manifold of spacetime satisfies wave equation,
\begin{eqnarray}
\Box R=0.
\end{eqnarray}
here $\Box=(\sqrt{-g})^{-1}\partial_{\mu}(\sqrt{-g}\partial^{\mu})$ is the d’Alembertian operator. 
If one passes to the linear regime, it is possible to probably observe gravitational waves in a natural form by adapting a suitable gauge frame. Note here that the general class of Lagrangians including the one without matter written above has been considered for theories of quantum gravity.

In the present work, we are going to investigate the cosmological implications of Nash theory by adding the matter fields. The modified action takes the following form:
\begin{eqnarray}
{\cal S}= -\frac{1}{2\kappa^2}\int\Big(2R^{\mu\nu}R_{\mu\nu}-R^2\Big)\sqrt{-g}d^{4}x+S_{\rm matter}\,, \label{action}
\end{eqnarray}
where $R_{\mu\nu}$ and $R$ are the Ricci tensor and Ricci (curvature) scalar, respectively, while $g$ is the determinant of the background metric tensor, $g_{\mu\nu}$ and $\kappa^{2}\equiv 8\pi G$. Here we have added the matter field sector, $S_{\rm matter}$, to the empty spacetime  action. The model given by an action (\ref{action}) with $S_{\rm matter}=0$ has been widely investigated in Ref. \cite{Aadne:2017oba}. As it has been known, the exterior vacuum metric is the Schwarzschild-(Anti) de Sitter (S(A)dS) instead of the Schwarzschild in GR. The reason is that in Nash gravity the highly nonlinear terms in the action (\ref{action}) lead to a second order differential equation rather than a first order one in the case of a static, spherically symmetric metric of the spacetime. As a result, the first integral of the field equation gives us a new integration constant. This new integration constant modifies the Schwarzschild  metric function $f(r)$  by a factor of $f(r)\propto -r^2$. This integration constant can be identified either as a positive (SdS) or negative (SAdS) case. This is very important because it distinguishes among the GR blackholes and Nash's versions. We can say that the Nash gravity blackholes and the GR ones have different asymptotic behaviors at $r\to\infty$. We have observed that the Nash blackholes have dS(AdS) boundary instead of a flat boundary in GR.

Using the above action, gravitational field equations are directly derived by taking into account the metric $g^{\mu\nu}$ as a dynamical field to yield
\begin{equation} 
\Box G_{\mu\nu} + G_{\alpha\beta}\Big(2R^{\alpha\beta}_{\mu\nu} - \frac{1}{2}g_{\mu\nu}R^{\alpha\beta}\Big)= \kappa ^2 T_{\mu\nu},\label{eq}
\end{equation}
where  $G_{\mu\nu}=R_{\mu\nu}-\frac{1}{2}g_{\mu\nu}R$ and  $T_{\mu\nu}=-\frac{2}{\sqrt{-g}}\frac{\delta S_{\rm matter}}{\delta g^{\mu\nu}}$. 
\par
Taking the trace of the equation of motion (EoM) given in Eq. (\ref{eq}), we obtain:
\begin{equation}
\Box R +\kappa^2 T=0
,\label{eq2}
\end{equation}
where $T\equiv T_{\mu}^{\mu}$ is the trace of the energy momentum tensor for the matter field action $S_{\rm matter}$. We assume that there exist a classical Einstein metric satisfying the vacuum field equation $G_{\mu\nu}=0$. Using Eq.(\ref{eq2}) we conclude that it satisfies this subclass model of $f(R,R_{\mu\nu}R^{\mu\nu})$ gravity in vacuum as well. Consequently any vacuum Einstein metric with $R=0$  is also a solution to Nash   gravity without matter contents\footnote{The reason is very clear, if $R=0$, then $\Box R=0$. If the matter sector contains any traceless matter field (like radiation), the above statement holds}. In this sense any asymptotic flat rotating metric (Kerr metric) is also solution to the divergence free  theory. In case of the cosmological constant it needs a sign reversion in the original action such that $2R^{\alpha}_{\,\,\mu}R^{\beta}_{\,\,\nu} \to -2R^{\alpha}_{\,\,\mu}R^{\beta}_{\,\,\nu} $ \cite{Lake:2017uic}.

Nevertheless, it should be noted that this theory contains higher-order time derivative in the equations of
motion. In a quantum version, higher-derivative interactions result (ghost) fields with negative norm yielding negative probabilities and possibly a breakdown in unitarity. In some cases, the higher-order time derivative in the equations of motion can be reduced to the second order in which such stability can be solved. The authors of Ref. \cite{Channuie:2018din} demonstrated that Nash theory possesses an instability in vacuum. However, in the present work, investigating classical aspects of divergence free theory with matter content in the cosmological background is our main objective.

\subsection{A flat FLRW geometry}
The geometry compatible with the
homogeneous and isotropic universe is the Friedmann-Lemaitre-Robertson-Walker (FLRW)
space-time with the corresponding line element:
\begin{eqnarray}
ds^2=-dt^2+a(t)^2\Big(\frac{dr^2}{1-kr^2}+r^2(d\theta^2+\sin^2\theta d\phi^2)
\Big),
\end{eqnarray}
 where we have expressed the spatial section in terms of spherical coordinates, $(r,\theta,\phi)$. The constant $k$ encodes the curvature of the space-time, with $k=0$ corresponding to flat
(Euclidean) spatial sections, and $k=\pm 1$ corresponding to positive and negative curvatures,
respectively. Here we define the Hubble parameter as $H\equiv \frac{d\ln a}{dt}$. The energy-momentum tensor of the matter contents is given in the following form:
\begin{eqnarray}
T_{\mu\nu}=(\sum_{i}\rho_i+\sum_i p_i)u_{\mu}u_{\nu}-(\sum_i p_i) g_{\mu\nu},\label{em}
\end{eqnarray}
where $i=\{\rm dark\,\,energy,\,dark\,\,matter, \,radiation\}:=\{\rm de,\,dm,\,r\}$. Here if we define a new parameter such that $\zeta=H^{1/2}$, we obtain equations of motion (EoMs) in the flat FLRW space-time:
\begin{eqnarray}
&&\frac{d}{dt}(\dot{\zeta}+\zeta^3)=\kappa^2\sum_{i}\rho_i \,,\label{eom1}
\\&&\zeta\frac{d^2}{dt^2}(\dot{\zeta}+\zeta^3)+\frac{3}{2}\frac{d}{dt}(\dot{\zeta}+\zeta^3)^2 =-\kappa^2(\sum_{i}\rho_i+\sum_i p_i)\label{eom2}.
\end{eqnarray}
It was found that the solutions of Eqs.(\ref{eom1}) and (\ref{eom2}) are investigated in Ref.\cite{Aadne:2017oba} for empty space. In the cosmological vacuum case when $T_{\mu\nu}\equiv 0$, in the absence of the matter fields, even the first Friedman equation for the scale factor turns out to be a second order instead of the first one in GR, 
\begin{eqnarray}
&&\frac{d}{dt}(\dot{\zeta}_{vac}+\zeta_{vac}^3)=0\,.\label{eom10}
\end{eqnarray}
The first integral of this differential equation gives us a new integration constant along with a new extra term which is proportional to the $H^{3/2}$,
\begin{eqnarray}
&&\dot \zeta_{vac}+\zeta_{vac}^3=C, \ \ C\in\mathcal{R}\,.\label{eom20}
\end{eqnarray}
The second-order field equation (\ref{eom2}) is solved trivially. The general solution for the Hubble $H$ in (\ref{eom20}) can be expressed in terms of the elementary functions,
\begin{eqnarray}
&&\frac{1}{3 C^{2/3}}\log\big(\frac{\sqrt{C^{2/3} + C^{1/3} \zeta_{vac} + \zeta_{vac}^2}}{C^{1/3} - \zeta_{vac}}
\big)
+\frac{\tan^{-1}\Big(\frac{( \zeta_{vac}C^{-1/3} + 1}{\sqrt{3}}
\Big)}{\sqrt{3} C^{2/3}} =t-t_0\,.
\end{eqnarray}
One can invert the expression and find exact $\zeta_{vac}(t)$. For $\zeta_{vac}\ll 1$, we have the following (Taylor series) expression for Hubble parameter,
\begin{eqnarray}
&&H_{vac}(t)\approx C^2(t-t_0-\frac{\pi}{6\sqrt{3}C^{2/3}})^2+\mathcal{O}(H_{vac}(t)^4)\,,
\end{eqnarray}
and for large enough Hubble values we have the following asymptotic solution (Puiseux series),
\begin{eqnarray}
H_{vac}(t)\propto (t-t_0)^{-1/2}+\mathcal{O}(H_{vac}(t)^{-4})\,.
\end{eqnarray}
As one can check Ref. \cite{Aadne:2017oba}, the non linearity of the Nash gravity drastically changes the form of the exact solutions in both cases of the blackhole or cosmological solution even when the matter contents are absent. It shows that Nash theory is not only different from GR at the action level but also predicts new solution totally different from GR ones. In the non vacuum case at radiation dominated epoch, the exact solutions for Nash's theory of  gravity are obtained widely in \cite{Lake:2017uic}. They showed that the solutions are quite different from the standard GR solutions.  However because we are not going to find exact solutions anymore, it makes it difficult to compare our mixture fluids model and their numerical (analytical) solutions with the important solutions obtained in Ref. \cite{Lake:2017uic}. Furthermore, it seems that Nash theory is more complicated than GR because of the existence of the higher-order derivative terms. Hence one doesn't need to include any matter contents to the action to  study late or early cosmology. We should clarify that the original Nash gravity without matter can't give late time acceleration expansion by itself. The reason is that it was shown in Refs.\cite{Aadne:2017oba,Lake:2017uic}, that one can't find a reasonable physically correct deceleration parameter or equation of state parameter guaranteed by the accelerating epoch of the current universe. To make it more clear, to have acceleration expansion one needs to either take $\ddot{a}>0$ or to have a negative deceleration parameter $q<-1$, being negative,
\begin{eqnarray}
\frac{dH_{vac}}{dt}=H_{vac}^2(1+q)\,,
\end{eqnarray}
which the observation data prefers ($q\sim -0.51$) and is equivalent to $\frac{dH}{dt}<0$. In the vacuum Nash gravity equation (\ref{eom20}) shows that 
\begin{eqnarray}
\frac{dH_{vac}}{dt}=2(C\sqrt{H_{vac}}-H_{vac}^2)\,.
\end{eqnarray}
The constraint is to have the accelerating universe, if $C>0$, when: 
\begin{eqnarray}
H_{vac}>C^{2/3}\,.
\end{eqnarray}
This constraint leads to $q<-1$ but it proposes a lower bound on the Hubble's constant\footnote{The algebraic equation for $\dot H_{vac}$ gives one negative $H$ as well as a positive constant $H$'s. The positive roots indicate deSitter solutions in the model }. It shows that the vacuum Nash gravity doesn't describe the current accelerating universe without including any matter content. That is the reason one needs to insert matter components. Here, in this work, we have included all types of the matter components to keep a track of the history of the Universe in the light of the current data.
\par
Let us go back to the Nash theory with matter components. By substituting Eq.(\ref{eom1}) into Eq.(\ref{eom2}), we find the following equation:
\begin{eqnarray}
&&\zeta\frac{d}{dt}\sum_{i}\rho_i+3\kappa^2\sum_{i}\sum_{j}\rho_i(t)\int_{t}\rho_j(t')dt'+\sum_i (\rho_i+p_i)=0, \label{continuty1}
\end{eqnarray}
which is the so-called "continuity equation’'. Note that in spite of the continuity equation in GR, this is a nonlocal equation in which the integration of $\rho$ is primarily required in order to quantify $\rho$. In the section below, we examine in more details this equation and aim to transform it into an ordinary differential equation instead of the integro-differential equation. 

\subsection{The collision term and effective energy density action}
Notice in Eq.(\ref{continuty1}) that there is a term in which $\rho_i$ couples with $\rho_j$, i.e., $\sum_{i}\sum_{j}\rho_i(t)\int_{t}\rho_j(t')dt'$. It will drastically influence the dynamics of the cosmological behaviors. We use a new re-parametrized time coordinate $\tau\equiv \int_t H^{-1/2}(t')dt'$ to rewrite  the continuity equation to yield 
\begin{eqnarray}
\frac{d}{d\tau}\sum_{i}\rho_i+3\kappa^2\sum_{i}\sum_{j}\rho_i(\tau)\int_{\tau}\rho_j(\tau')\sqrt{H(\tau')}d\tau'+\sum_i (\rho_i+p_i)=0.
\label{continuty2}
\end{eqnarray}
Supposing $p_{\rm tot}=p(\rho_{\rm tot})$, we can rewrite the above continuity equation in terms of the total energy density to obtain: 
\begin{eqnarray}
\frac{d}{d\tau}\rho_{\rm tot}+3\kappa^2\rho_{\rm tot}(\tau)\int_{\tau}\rho_{\rm tot}(\tau')\sqrt{H(\tau')}d\tau'+ (\rho_{\rm tot}+p(\rho_{\rm tot}))=0\label{continuty-tau}.
\end{eqnarray}
In the equivalent form ($\rho_{\rm tot}\equiv \rho,p(\rho)\equiv f(\rho))$, it turns out to yield the following differential equation:
\begin{eqnarray}
\frac{d}{d\tau}\Big[\frac{d}{d\tau}\ln\rho+\frac{f(\rho)}{\rho}
\Big]+3\kappa^2\rho\sqrt{H(\tau)}=0\label{rhot}.
\end{eqnarray}
Notice that Eq.(\ref{rhot}) is a nonlinear second order differential equation for the energy density $\rho$ in which the solution depends on the time evolution of the cosmological background depending on $H(\tau)$. There is no any simple way to figure out the energy density profile except using a reconstruction scheme. Here we will assume a form for the Hubble parameter allowing to integrate Eq.(\ref{rhot}) analytically. Let us compare new the continuity equation given in Eq.(\ref{rhot}) with the one in Einstein relativity such that 
\begin{eqnarray}
\frac{d}{d\tau}\ln \rho+3H^{3/2}\left(1
+\frac{f(\rho)}{\rho}\right)=0. \label{rhoGR}
\end{eqnarray}
It is clear that  we use the same fluid model as $p=f(\rho)$ in the Einstein gravity to compare it with our $f(R,R_{\mu\nu}R^{\mu\nu})$ case. It is worth mentioning here that the continuity equation Eq.(\ref{rhot}) is a second-order differential equation instead of the first order one. The reason is that $f(R,R_{\mu\nu}R^{\mu\nu})$ gravity is constructed from the scalar curvature squared terms instead of the Ricci scalar. As a result, the continuity equation resulting from the Bianchi identity applied to the left-hand side of the modified Einstein field equation leads to a second order differential equation. 

It is intuitive to derive an effective action for the matter content density in which the density function $\rho$ governs the EoM given in Eq.(\ref{rhot}). Referring to the Euler-Lagrange equation, we conclude from Eq.(\ref{rhot}) that there should be an effective conjugate momentum $p_{\rho}$ corresponding to the density function $\rho$ which is derived from an effective Lagrangian $\mathcal{L}_{\rho}$ such that
\begin{eqnarray}
p_{
\rho}\equiv \frac{\partial\mathcal{L}_{\rho}}{\partial \dot\rho}
=
\frac{d}{d\tau}\ln\rho+\frac{f(\rho)}{\rho}\label{rrho}.
\end{eqnarray}
A possible potential function can be deduced to obtain
\begin{eqnarray}
V(\rho)\equiv -3\kappa^2\int_{\rho}\rho'\sqrt{H(\rho')}d\rho' \,.\label{Vrho}
\end{eqnarray}
In order to explicitly quantify the potential we need to specify the functional form of the Hubble parameter and energy density. Hence, we propose the following irregular effective action of the field $\rho$,\footnote{This form of effective action is selected among a list of possibilities and there are more non-canonical forms for the effective action in holonomic (explicitly time independent) gauge. There are other possible action forms where the action can be constructed from time dependent " Lennard-Jones potential " terms like $V\left[\rho(\tau),\dot\rho (\tau);\tau \right]$. }
\begin{eqnarray}
S_{\rm eff}\left[\rho(\tau),\dot\rho (\tau) \right]=\int d\tau \dot \rho \Big(\frac{\dot\rho+f(\rho)}{\rho}-V(\rho)
\Big).
\end{eqnarray}
Varying the above effective action with respect to the energy density $\rho$ gives us the irregular continuity equation present in Eq.(\ref{rhot}). We expect that there are some physically important results related to the effective action given above. However, we will leave these interesting for our future work.

\subsection{De Sitter like  expansion and energy density profile }
\begin{figure}[H]
\begin{center}
	\includegraphics[width=.6\textwidth]{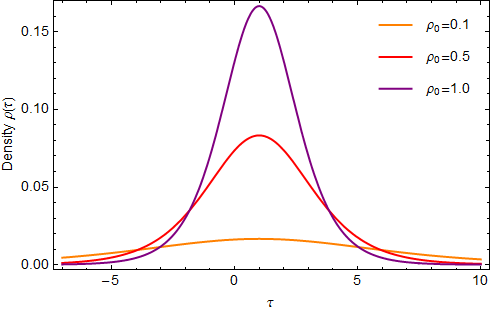}
	\caption{The plot shows the time evolution of the energy density $\rho(\tau)$ with three different values of $\rho_{0}$. Notice that at late time the energy energy tends to zero. The acceptable scenario here is a viable energy content matter in the Universe initiated from a low dense matter at very early epochs and then the density slowly increases to a peak of energy density where the universe reaches a turning point with the maximum energy content. Note that under this circumstances the system passed an unstable phase transition point. The final state of the Universe is dominated by an energy density of order of a non fine-tuned cosmological constant $\Lambda_{\rm non-Fin}$. The reason for absence of a fine tuning cosmological constant is the potentially renormalized form of the  $f(R,R_{\mu\nu}R^{\mu\nu})$ gravity as an alternative model for quantum gravity.    }
	\label{rho}
  \end{center}
\end{figure}

It is illustrative to show how the solution looks like when the Universe is dominated by barotropic fluid $p=w\rho$ and when $H(\tau)\approx H_0$ is de sitter like epoch. In this general case, we obtain the following solution for vacuum energy density from Eq. (\ref{rhot}) ,
\begin{eqnarray}
&&\rho(\tau)=\frac{\rho_0}{6 \kappa ^2 \sqrt{H_0} }\cosh ^{-2}\left(\frac{ \sqrt{\rho_0}}{2} \left( \tau-\tau_0\right)\right)\label{2ndrho}.
\end{eqnarray}
Here $\rho_0, \tau_0, H_0$ are integration constants and we set $H_0=1$. In Fig.(\ref{rho}), we plot this barotropic energy density for different values of the present energy density, $\rho_{0}$. Note that in late time when $\tau\to\infty$ we find $\rho(\tau)\to 0$ which means that the Universe has an empty energy density. Remarkably, we observe that the total energy density both in early and late time Universe vanishes and the whole cosmological background undergoes a de Sitter expansion from vacuum to a future vacuum cosmological constant dominated epoch. 

Although the total energy density vanishes at early (and late) times, there is no need to have a pure vacuum. Let us clarify this very carefully. At very early time, the universe has radiation as dominant energy density which in a classical Einstein-Hilbert action is governed by $\rho_r\sim a(t)^{-4}$. In our cosmological model, we need to assume that there are other types of matter contents. For example, we have a type of primary tachyonic field  $\psi$ with  energy density given by $\rho_{\psi}\sim-\rho_r$ in which this equal sized energy content is balanced with radiation field maintaining the vanishing total energy density at time $\tau\to -\infty$. However, this possible form of energy content violates the null energy condition. We can suppose that it can be generated by a wormhole source at early universe. As a result we can predict the existence of a primary wormhole in a similar form as primary black holes. Unfortunately we are not able to probe the form of this type of the tachyonic matter. Since the total pressure is governed by a barotropic equation of state, we can just conclude that the tachyonic matter is also a barotropic exotic fluid.
\subsection{Universe filled with barotropic fluid with energy density $H\propto \rho^{-2}$}
\begin{figure}[H]
\begin{center}
	\includegraphics[width=.6\textwidth]{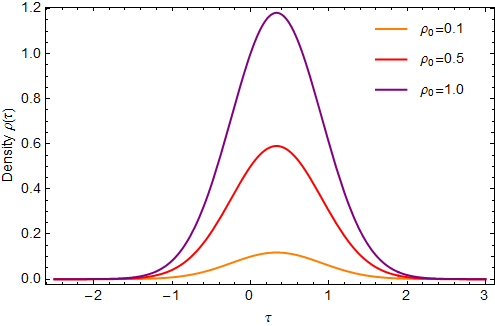}
	\caption{The plot shows the time evolution of the energy density $\rho(\tau)$ with three different values of $\rho_{0}$. Notice that at late time the energy energy tends to zero. In this scenario the universe initiated from an almost de-Sitter epoch , where the dominated energy density was very low and after a time interval, the universe reaches the maximum amount of the energy at a turning point, where the phase changed from deceleration to the acceleration. At the late time the dominant energy content is cosmological constant and the energy density decreases. Note that in this scenario the cosmological constant is not fine tuned. There is an effective cosmological constant which mimics the background cosmological evolution. An existence of a non fine tuned unique cosmological constant is due to the renormalizable quantum version of $f(R,R_{\mu\nu}R^{\mu\nu})$ gravity and the appearance of higher order curvature terms. }
	\label{rhoD1}
  \end{center}
\end{figure}

The continuity equation of energy density given in Eq.(\ref{rhot}) is  nonlinear and highly coupled to the Hubble parameter \textit{H}. To make it integrable, we should not only know the initial energy density profile at very early times, but also we need to know how Hubble parameter evolves with respect to the density. An easy case and physically remarkable situation is when we suppose that  $H= \frac{H_0}{\rho^2}$. Using this assumption we can integrate Eq.(\ref{rhot}) and find \footnote{Note that here $H_0$ is an arbitrary constant and it may related to the initial Hubble parameter $H(\tau=0)$ as well as initial energy density $\rho(\tau=0)$.}:
\begin{eqnarray}
\rho(\tau)=\rho_0\exp\left(\frac{\tau}{\tau_0}-\frac{3\kappa^2\sqrt{H_0}\tau^2}{2}\right)
\label{3ndrho}.
\end{eqnarray}
This is a Gaussian density profile plotted in Fig.(\ref{rhoD1}). We observe that for very early times, when $\tau \to -\infty, \rho(\tau) \to 0$ as well as late time when $\tau \to \infty$. Interestingly, we also observe a maximum of energy density at a certain time. Note that this form of energy density has a same physics as the one obtained in Eq.(\ref{2ndrho}). Although in this model the universe didn't evolve exponentially, and the total Hubble scale factor is not de Sitter, still, the Universe will start from a vacuum at very early times and comes to an end to vacuum. 

In this scenario the evolution form of the universe is Gaussian, and only at an instant of time namely $\tau_m$, it reaches the  maximum of energy density and hence the minimum of the Hubble scale. This time scale corresponds to an unstable point in the cosmic time, implying that the energy density reaches its maximum as an unstable transition point. Then the system suddenly undergoes a very sharp phase transition from maximum energy content to the vacuum. In our new scenario, the Universe starts from de Sitter (supporting the idea of de Sitter inflationary era), followed by the matter rich content epoch as an unstable time, and then transits to the late time cosmological constant dominated epoch: 
\begin{eqnarray}
\tau_m=\left(3\kappa^2\sqrt{H_0}\tau_0\right)^{-1},\ \ \rho_m=\rho_0\exp\left(
\frac{1}{6 \sqrt{H_0}\tau_0^2 \kappa ^2}\right).
\end{eqnarray}
Now we can find the Hubble parameter written in terms of the cosmic time $\tau$:
\begin{eqnarray}
H(\tau)=\frac{H_0}{\rho_0^2}\exp\left(-\frac{2\tau}{\tau_0}+3\kappa^2\sqrt{H_0}\tau^2\right)
\label{H1}.
\end{eqnarray}
We will compare the predictions of this model with observational data in section (\ref{om}). For this purpose, it is adequate to rewrite it in terms of the redshift $1+z=\frac{1}{a(\tau)}$. Here we suppose that $a(\tau=0)=a_0\equiv 1$ for simplicity, where $a(\tau)=\exp\{\int_{\tau}H(\tau')d\tau'\}$. Using an integration of the exponential function: 
\begin{eqnarray}
\int\exp(-a x+bx^2)dx=\frac{\sqrt{\pi } H_0 e^{-\frac{a^2}{4 b}} \text{erfi}\left(\frac{2 b x-a}{2 \sqrt{b}}\right)}{2 \sqrt{b} \rho_0^2},
\end{eqnarray}
where the "imaginary error function" defined by $\text{erfi}(y)=-i\,\text{erf}(iy)$, we obtain:
\begin{eqnarray}
a(\tau)=\exp\left(\frac{\sqrt{\pi } H_0 e^{-\frac{a^2}{4 b}} \text{erfi}\left(\frac{2 b \tau-a}{2 \sqrt{b}}\right)}{2 \sqrt{b} \rho_0^2}
\right),\ \ a\equiv \frac{2}{\tau_0},\ \ b\equiv 3\kappa^2\sqrt{H_0}
\label{a1}.
\end{eqnarray}
Consequently, we can write the redshift $z$ in terms of time $\tau$ as:
\begin{eqnarray}
1+z=\exp\left(-\frac{\sqrt{\pi } H_0 e^{-\frac{a^2}{4 b}} \text{erfi}\left(\frac{2 b \tau-a}{2 \sqrt{b}}\right)}{2 \sqrt{b} \rho_0^2}
\right)
\label{z1}.
\end{eqnarray}
Using the above expression, we obtain $\tau$ written in terms of the redshift $z$: 
\begin{eqnarray}
\tau=\frac{\left(a+2 \sqrt{b}\right)\pi H_{0}}{8 b \left(\sqrt{b} \rho_0^2 e^{\frac{a^2}{4 b}} \log (z+1)\right) \, _1F_1\left[\frac{1}{2};\frac{3}{2};\left(\frac{2 \sqrt{b} e^{\frac{a^2}{4 b}} \rho_0^2 \log (z+1)}{H_0 \sqrt{\pi }}\right)^2\right]}.\label{zt}
\end{eqnarray}
Here we have used the relation $\text{erfi}(z)=-i\,\text{erf}(iz)$ and $\text{erf}(z)=(2z/\sqrt{\pi})_{1}F_{1}[1/2,3/2,-z^{2}]$ with $_{1}F_{1}[1/2,3/2,-z^{2}]$ being the first kind of the hypergeometric function. Using $\tau$ given in Eq.(\ref{zt}) we can express the Hubble parameter in terms of the redshift, $H=H(z)$. Therefore, we write
\begin{eqnarray}
H(z)=\frac{H_0}{\rho_0^2}\exp\left(-\frac{2\tau(z)}{\tau_0}+3\kappa^2\sqrt{H_0}\tau(z)^2\right).
\label{H-model1}
\end{eqnarray}
\begin{figure}[H]
\begin{center}
	\includegraphics[width=.47\textwidth]{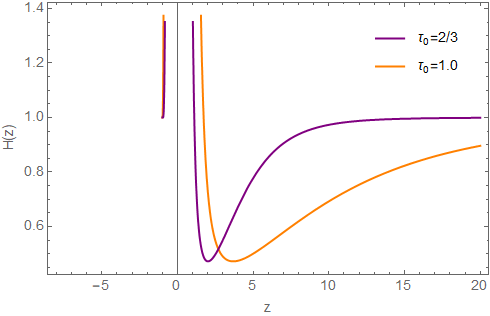}
    \includegraphics[width=.47\textwidth]{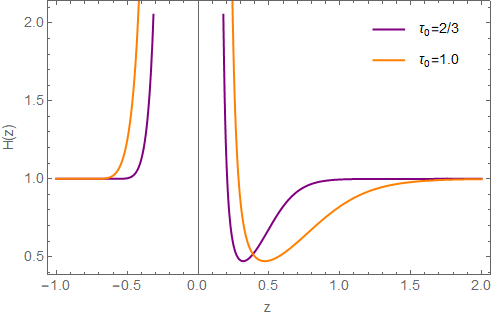}
	\caption{The plots show the Hubble parameter $H(\tau)$ given in Eq. (\ref{H-model1}) versus redshift $z$ explicitly expressed in Eq. (\ref{zt}) using $\rho_{0}=0.5$ (left panel) and $\rho_{0}=1.0$ (right panel).}
	\label{Hz}
  \end{center}
\end{figure}
The variations of the Hubble parameter with the redshift are displayed in Fig.(\ref{Hz}). We notice for Fig.(\ref{Hz}) that in very early time, i.e. $z\gg 1$, the parameter $H(\tau)$ is an increasing function of $\tau$ and near the big bang time it tends to a constant, in support of the idea of inflationary universe. At the present time near $z\sim0$, the Hubble becomes very large and the phases is completely divided specially when $z<0$ and the universe undergoes a deceleration phase. As a result our model predicts a type of phase transition from deceleration to acceleration.

\subsection{Exact Nash cosmology for almost uniform total energy density}

Let us first start by considering Eq.(\ref{eom1}) and figuring out its solution. Suppose that an arbitrary form of total energy density is given by $\rho_{\rm{tot}}(t)\equiv \sum_{i}\rho_i$. Using Eq.(\ref{eom1}) we obtain:
\begin{eqnarray}
&&\dot{\zeta}+\zeta^3-\kappa^2\int_t\rho_{\rm{tot}}(t')dt'=0.
\end{eqnarray}
Using a standard method, we can simply solve the above nonlinear equation by changing a variable such that $\zeta\to \psi=\zeta^{-2}$, and the exact solution for the above matter contents when $\int_t\rho_{\rm{tot}}(t')dt'\approx {\rm constant}=\rho_0>0$ takes the form:
\begin{eqnarray}
\frac{1}{6}\log{\left({{\zeta(t)}^{2}}+{{\rho_0}^{\frac{1}{3}}}\,\zeta(t)+{{\rho_0}^{\frac{2}{3}}}\right) }+\frac{\sqrt{3}}{3}\arctan\left( \frac{2\zeta(t)+{{\rho_0}^{\frac{1}{3}}}}{\sqrt{3}\,{{\rho_0}^{\frac{1}{3}}}}\right) -\frac{1}{3}\,\log{\left(\zeta(t)-{{\rho_0}^{\frac{1}{3}}}\right) }={\rho_0}^{\frac{2}{3}}\left(t-t_0\right).\label{sozeta}
\end{eqnarray}
Apart from the condition $\int_t\rho_{\rm{tot}}(t')dt'\approx {\rm constant}=\rho_0>0$, we can not figure out any exact solution of the above nonlinear differential equation.
\begin{figure}[H]
\begin{center}
	\includegraphics[width=.6\textwidth]{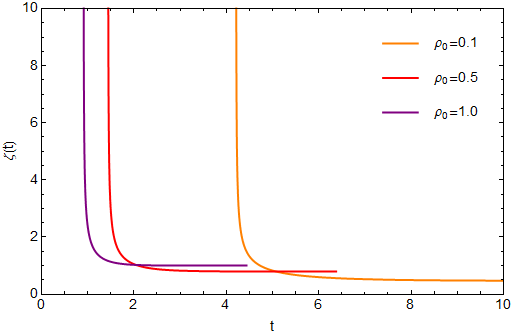}
	\caption{The plot shows the time evolution of the solution $\zeta(t)$ given in Eq.(\ref{sozeta}).}
	\label{zeta}
  \end{center}
\end{figure}
The time-evolution of the solution $\zeta(t)$ given in Eq.(\ref{sozeta}) is illustrated in Fig.(\ref{zeta}). It can be attributed to the behavior of the Hubble parameter since $\zeta=H^{1/2}$. Notice that the Hubble parameters are decreasing functions and become constant in the present epoch in all cases. This is an exact solution for the cosmological EoM and as a dominant solution in the early epoch can be used to build up a quasi stable de Sitter solution. 

\subsection{A barotropic fluid solutions}

In the previous subsection, we can only obtain the exact solution for $\zeta$ in the case of uniform total matter density. However, it is also possible to figure out exact solutions if we use the scale factor representation of this equation. Specially using the reconstruction technique we first suppose that $\rho_{\rm tot}={\tilde \rho}_{0} a^{-3 (1+w)}$ with ${\tilde \rho}_{0}=\rho(a_{0}=1)$. This is a standard form of the energy density written in terms of the scale factor. Here $w$ is the equation of state parameter (EoS). In the barotrooic fluid manner, the EoS is parametrized by $p=w\rho$ responsible for a time evolution. Therefore, the EoM in Eq.(\ref{eom1}) can be recast as follows:
\begin{eqnarray}
\frac{\ddot{a}}{a}+\left(\frac{\dot{a}}{a}\right)^2=2{\tilde \rho}_{0} a^{-3(w+\frac{1}{2})}\dot{a}^{1/2}.
\label{a-eom1}
\end{eqnarray}
The above equation can be analytically solved to obtain
\begin{eqnarray}
{\tilde \omega}(t-t_0)=a^{5/2} \left(\frac{c_1}{a^{3/2}}+\frac{3 {\tilde \rho}_{0}  a^{\frac{1}{2}-3 w}}{2-3 w}\right) \,
   _2F_1\left(1,a_1;1+b\,|\,c\right),
\end{eqnarray}
where $c_1$ is a constant and $_{2}F_{1}(1,a_1;b\,|\,c)$ is the hyper-geometric function with $a_1=(8-3 w)/(6-9 w),\,b=2/(2-3 w)$ and $c=(3 a^{2-3 w} {\tilde \rho}_{0})/(3 w-2) c_1$. Furthermore, $t_0$ denotes the initial time, ${\tilde \omega}$ is an integration constant which can be interpreted as a frequency of oscillation. The exact solution of a scale factor for radiation dominated Universe with $w=\frac{1}{3}$ was studied in \cite{Aadne:2017oba}. In our generalized solution for the barotropic fluid, if one uses $w=\frac{1}{3}$, the scale factor is simplified to the following form  
\begin{eqnarray}
{\tilde \omega}(t-t_0)=\left(c_1 a+3 {\tilde \rho}_{0}  a^{2}\right) \,
   _2F_1\left(1,\frac{7}{3};3\,|\,-3a {\tilde \rho}_{0}c_1\right),
\end{eqnarray}
It is possible to invert this expression and find $a(t)$.

\subsection{Perturbation analysis of energy density via Hartman-Grobman  \&  Lyapunov  linearizion theorems }

In any linear system, we classify physical behaviors of any fixed point by using the eigenvalues of the matrix constructed from a $n$-dimensional differential equation. However, the situation has dramatically changed when working in the nonlinear one. In the latter, the behavior of the system is more difficult to handle. Fortunately, we can transform the nonlinear differential system to a linear one. In doing so, we find the Jacobian matrix, ${\mathcal J}$, corresponding to the system and evaluate it at the fixed point. As a result, regarding the Hartman-Grobman theorem, we obtain a linear system with a characteristic coefficient matrix. Let's consider a differential equation (DE): $\vec{X}'=d\vec{X}/dN=f(\vec{X})$ defined on ${\mathbb R}^{n}$, where $N$ plays the role of time and $\vec{X}$ is a vector field. If ${\vec a}$ is an equilibrium point $f({\vec a})=0$, the linear approximation of $f(\vec{X})$ at ${\vec a}$ yields
\begin{equation}\label{ds011}
f(\vec{X})\approx {\mathbb D}f(\vec{a})(\vec{X}-\vec{a})\,,
\end{equation}
where
\begin{equation}\label{ds101}
({\mathcal J})_{ij}\equiv{\mathbb D}f(\vec{a})=\left(\frac{df_{i}}{dX_{j}}\right)_{\vec{X}=\vec{a}}\,,
\end{equation}
is the derivative (Jacobian) metrix of $f$. Therefore, with the given $\vec{X}'=f(\vec{X})$, we associate the linear DE using
\begin{equation}\label{ds110}
\vec{U}'= {\mathbb D}f(\vec{a})\vec{U}\,,
\end{equation}
where $\vec{U}=\vec{X}-\vec{a}$, called the linearization of the DE at the equilibrium point $\vec{a}$. It is worth noting that the solutions of Eq.(\ref{ds110}) will approximate the solutions of the nonlinear DE in a neighborhood of the equilibrium point $\vec{a}$ provided that the equilibrium point is hyperbolic. This means that all eigenvalues ($\lambda_{i}$) of ${\mathbb D}f(\vec{a})$ have non-zero real part, $\Re e(\lambda_{i})\neq 0$.

Now we are going to quantify the stability of the fixed points obtained from Eq.(\ref{rhot}) by using the above linearization. Consider the EoM for the total energy density given in Eq.(\ref{rhot}). We then transform it to the second-order differential equation to obtain
\begin{eqnarray}
{\ddot \rho}-\frac{{\dot\rho}^2}{\rho}-\Big(\frac{f(\rho)}{\rho}-f'(\rho)\Big){\dot\rho}+3\kappa^2\rho^2\sqrt{H(\tau)}=0\,, \label{eqrho}
\end{eqnarray}
where the dots denote derivatives with respect to $\tau$. Specifically, the above DE is nonlinear. In order to figure out its solutions, we use the following change of the variables:
\begin{eqnarray}
&&\dot{\rho}=X\label{dyn-eq1}\\
&&\dot{X}=X\Big(\frac{f(\rho)}{\rho}-f'(\rho)\Big)-3
   \kappa ^2 \rho ^2\sqrt{H(\tau)}+\frac{X^2}{\rho}\label{dyn-eq2}.
\end{eqnarray}
Here $f'(\rho)=\frac{df(\rho)}{d\rho}$. This in a non autonomous dynamical system describing a continuous-time nonlinear density $\rho$. In the language of the dynamical systems, $\rho(\tau)$ is the {\it state} of the system and we can treat the Hubble $H(\tau)$ as the control input. Note that the left hand side of the  Eq. (\ref{dyn-eq2}) is a {\it Lipschitz} or continuously  differentiable nonlinear function. A standard way to study the time evolution of the density function is to investigate the trajectory $\phi_t(\rho_0)$ where $\rho_0\equiv \rho(\tau=0)$ is the initial density profile.

Using the Hartman-Grobman linearizion theorem, we find that the fixed points for the system are located at $P=(X_c=0,\rho_c=\epsilon(\to 0))$, and the corresponding Jacobian matrix reads
\begin{eqnarray}
(\mathcal{J})=
 \begin{bmatrix}
  \frac{f(\epsilon )}{\epsilon }-f'(\epsilon ) & -6 \epsilon\kappa^2\sqrt{H(\tau_
0)}  \\
 1 & 0 
 \end{bmatrix}.
 \end{eqnarray}
To ensure the stability of the solutions, it is enough to set all eigenvalues of the Jacobian matrix so that $\lambda_i$ satisfies $\mathrm{Re}(\lambda_i)\neq 0$. Therefore in our case, we find
\begin{equation}
\lambda_{\pm}=\frac{-\epsilon  f'(\epsilon )+f(\epsilon )\pm\sqrt{\left(\epsilon  f'(\epsilon )-f(\epsilon )\right)^2-24 \kappa ^2 \epsilon
   ^3\sqrt{H(\tau_
0)}}}{2 \epsilon }.
\end{equation}
Note that when $\epsilon\to 0$ we have (if $f(0)>0$):
\begin{equation}
\lambda_{\pm}=\frac{\epsilon^2}{6|f(0)|}\Big(36\kappa^2\sqrt{H(\tau_
0)}+f(0)f'''(0)\pm|f(0)|f'''(0)\Big)\to 0.
\end{equation}
So the system is unstable under density perturbations. \par 
Moreover, there is the following important alternative  theorem to study stability of the above non-autonomous system: Lyapunov theorem. Regarding the Lyapunov theorem for nonautonomous dynamical systems \cite{Lyapunov}, we see that the system of Eqs.(\ref{dyn-eq1},\ref{dyn-eq2}) is global and over the entire connected domain $\mathcal{D}$, uniformly over the entire time interval $\Big[\tau_0, \infty\Big)$, and asymptotically stable about its equilibrium 
$P=(X_c,\rho_c)$, if there exist a Lyapunov function $V(X,\rho,\tau):\mathcal{D}\times \Big[\tau_0, \infty\Big)\to \mathcal{R} $ and three functions $\alpha,\beta,\gamma$ satisfying the following conditions:

\begin{itemize}
\item a: $V(X_c,\rho_c,\tau_0)=0$,
\item b: $V(X\neq X_c,\rho\neq \rho_c,\tau>\tau_0)>0$,
\item c: $ \alpha\sqrt{X^2+\rho^2}\leq  V(X,\rho,\tau)\leq \beta \sqrt{X^2+\rho^2}$, and
\item d: $\frac{d}{d\tau}V(X,\rho,\tau\geq \tau_0)\leq -\gamma \sqrt{X^2+\rho^2}<0$.
\end{itemize}
Now we need to quantify a suitable form for $V$. Note that from the above conditions, we find
\begin{eqnarray}
\frac{d}{d\tau}V(X,\rho,\tau\geq \tau_0)=X\frac{\partial V}{\partial \rho}+\frac{\partial V}{\partial X}\Big(X\Big(\frac{f(\rho)}{\rho}-f'(\rho)\Big)-3
   \kappa ^2 \rho ^2\sqrt{H(\tau)}+\frac{X^2}{\rho}\Big).
\end{eqnarray}
A suitable Lyapunov function  for a set of parameters $\{ \alpha,\beta,\gamma\}$ is given as the following:
\begin{eqnarray}
V(X,\rho,\tau)=V_0\Big[\exp\Big(-\alpha X^2-\beta \rho^2-\gamma (\tau-\tau_0)^2 \sqrt{X^2+\rho^2}
\Big)-1\Big].
\end{eqnarray}
Notice that it violates condition (d) at point $P=(X_c=0,\rho_c=\epsilon(\to 0))$. As a result the density equation is globally, uniformly and asymptotically unstable. This instability ensures the cosmological phase transitions during cosmic epochs.  

\subsection{Poincare portrait for ($\rho,\dot{\rho})$ and energy conditions}
\begin{figure}[H]
\begin{center}
	\includegraphics[width=.45\textwidth]{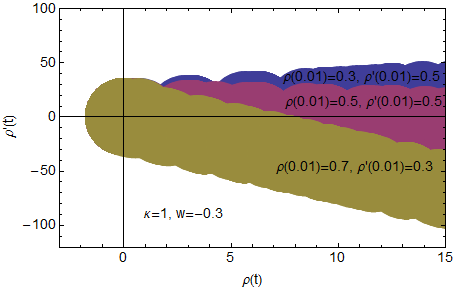}
    \includegraphics[width=.45\textwidth]{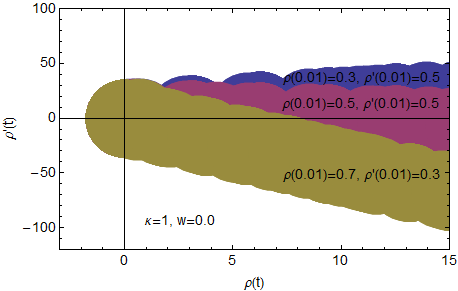}
    \includegraphics[width=.45\textwidth]{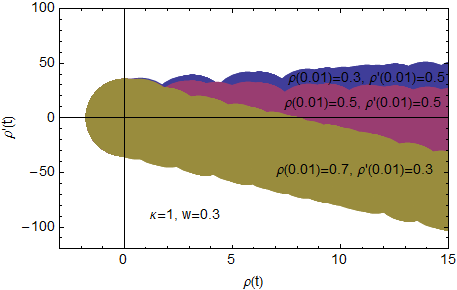}
    \includegraphics[width=.45\textwidth]{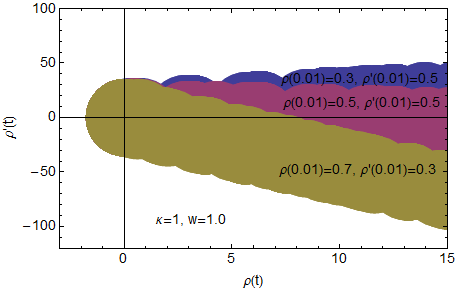}
	\caption{The plots show the phase portrait for pair $\rho-\dot\rho$ based on the generalized continuty equation given in Eq.(\ref{rhot}). From energy conditions point of view we conclude that there are regions of the cosmological epochs where the NEC is violated and they significantly indicate the existence of primordial transversal wormholes. Note that shaded regions above those of $\rho(0.01)=\dot{\rho}(0.01)=0.5$ display $\rho(\tau_0)<\dot{\rho}(\tau_0)$.}
	\label{rhorhop}
  \end{center}
\end{figure}

In this subsection, we aim to quantify the phase of ($\rho,\dot{\rho})$ and examine the relation between continuity equation and energy conditions. We also explain the possible phase of the matter in $f(R,R_{\mu\nu}R^{\mu\nu})$ gravitational theory. In so doing, we first explain the main idea of the continuity equation which obeys Eq.(\ref{rhoGR}). Having assumed that the matter is isotropic, it is worth noting that the energy-momentum tensor components, energy density $\rho$ and pressure $p$ should satisfy the following well known energy conditions \cite{Hawking} summarized in Table\,(\ref{energy}): 
\begin{table}[H]
\centering
\caption{In terms of
principal pressures, we provide four types of energy conditions. These energy conditions can be formulated via physical and effective approaches.}
\begin{ruledtabular}
\begin{tabular}{lll}
Energy condition & Physical approach & Effective approach \\
\hline	
null energy condition (NEC) & $T_{\mu\nu}k^{\mu}k^{\nu} \geq 0$ \& $k^{\mu}$ being a null-like vector & $\rho+p\geq 0$ \\
weak energy condition (WEC) & $T_{\mu\nu}U^{\mu}U^{\nu}\geq 0$ \& $U^{\mu}$ being a time-like vector & $\rho \geq 0, \rho+p \geq 0$ \\
dominant energy condition (DEC) & $T_{\mu\nu}U^{\mu}U^{\nu}\geq 0$, \& $T_{\mu\nu}U^{\mu}$ is not space-like & $\rho\geq 0,\, \rho  \pm p \geq 0$ \\
strong energy condition (SEC) & $(T_{\mu\nu}-g_{\mu\nu}T/2)U^{\mu}U^{\nu}\geq 0$ \& $T={\rm Tr}T_{\mu\nu}$ & $\rho  +p \geq 0, \rho +3p\geq 0$ \\
\end{tabular}
\end{ruledtabular}
\label{energy}
\end{table}
There are several different ways to formulate all the energy conditions deployed in classical general relativity. At least, we have three types of formulating them, i.e., the
geometric, the physical and the effective approach, see for example Ref.\cite{Curiel} and references therein. It was found in Ref.\cite{Visser:1999de} that the energy conditions are defined using a physical approach and are widely used in many different types of modified gravity theories, e.g., $f(R)$ theory \cite{fR}, $f(T)$ theory \cite{fT}, Gauss-Bonet theory \cite{GB} and other interesting models. Note here that the violation of the NEC violates all other energy conditions. Regarding the continuity equation in GR, if the WEC holds, then the only possibility is that $H<0$ meaning that the cosmological background is in the deceleration phase.  Furthermore, if we have NEC, then for an accelerating universe  it implies that $d\rho/d\tau<0$. In addition, if DEC holds, then the only possibility is to have a decelerating behavior.  In  the traversable wormhole scenarios, the NEC is violated. If we require a wormhole to be located in the accelerating universe, we need to have $d\rho/d\tau<0$. In our $f(R,R_{\mu\nu}R^{\mu\nu})$ gravity case, note that if $d\rho/d\tau<0$, from Eq.(\ref{rhot}) and if $\rho>0$, necessarily we will have 
\begin{eqnarray}
\frac{d^2}{d\tau^2}\rho<0.
\end{eqnarray}
The above inequality with the condition $\rho>0$ implies that $\frac{d}{d\tau}\rho<\dot{\rho}(\tau_0)$. Consider the phase portraits of $\rho-\dot\rho$ displayed in Fig.(\ref{rhorhop}). In our case, we need to consider $\dot{\rho}(\tau_0)>0$ to have a compact phase space. From the Figs.(\ref{rhorhop}), we observe that there are some regions where $\frac{d}{d\tau}\rho<\dot{\rho}(\tau_0)$ and $\rho>0$. Notice that shaded regions above those of $\rho(0.01)=\dot{\rho}(0.01)=0.5$ display $\rho(\tau_0)<\dot{\rho}(\tau_0)$. It demonstrates that the violation of the NEC is possible. Since this violation is detected for different types of the EoS $w$, for a barotropic fluid, we claim that a class of  the traversable wormholes probably exist in the primordial epochs of our Nash cosmological model. 

\section{Om Diagnostic Analysis}
\label{om}
Commonly the cosmological parameters like the Hubble parameter $H$, deceleration parameter $q$, and the equation of state (EoS) parameter
$\omega$ are important when checking the consistency of a particular model. However, $H$ and $q$ are not adequate in differentiating among dark energy
models. This is so since any dark energy
models can generate a positive Hubble parameter and a negative deceleration parameter, i.e., $H > 0$ and $q < 0$, for the present cosmological epoch. Therefore, higher-order time derivatives of the scale factor $a(t)$ is required to analyse the dark energy models \cite{alam,sah}.

Let's consider the first parameter, called the deceleration parameter $q(t)$, which is defined in terms of the Hubble parameter via:
\begin{eqnarray}
q(t)\equiv -\left(1+\frac{\dot H}{H}\right)\longrightarrow q(\tau)\equiv -\left(1+H^{-3/2}\frac{dH}{d\tau}\right),
\end{eqnarray}
where in order to transform $q(t)$ into $q(\tau)$, we have used $dH/dt=H^{-1/2}dH/d\tau$. Regarding our solution given in Eq.(\ref{H1}), we obtain from the above relation:
\begin{eqnarray}
q(\tau) = -1 +\Big(\frac{2-3 \sqrt{H_{0}} \kappa ^2 \tau_{0}}{\frac{\tau_{0}}{\rho_{0}}H_{0}^{1/2} }\Big)e^{\frac{-3 \sqrt{H_{0}} \kappa ^2}{2}  \tau+\frac{ \tau}{\tau_{0}}}.\label{42}
\end{eqnarray}
\begin{figure}[H]
\begin{center}
	\includegraphics[width=.47\textwidth]{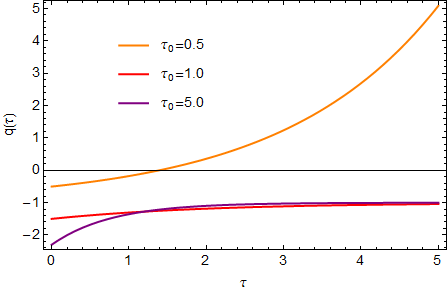}
\includegraphics[width=.47\textwidth]{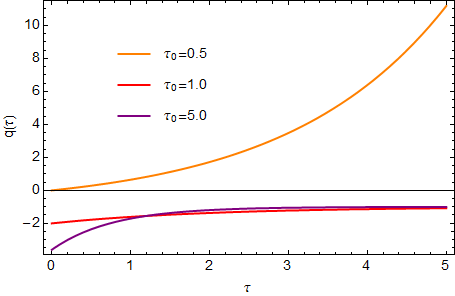}
	\caption{The plots show the evolution of the deceleration parameter $q(\tau)$ for different values of $\rho_{0}$ and $\tau_{0}$. We used on the left panel $\rho_{0}=0.5$ and on the right panel $\rho_{0}=1.0$. We used $\kappa=1$ and $H_{0}=1$ for the plots. The left panel using $\rho_{0}=0.5$ displays an accelerating expansion for all time with $\tau_{0}=1.0$ and $5.0$. However, we
have another case of phase transition from an acceleration to a deceleration with $\tau_{0}=0.5$. The right panel utilizing $\rho_{0}=1.0$ displays an accelerating
expansion for all time with $\tau_{0}=1.0$ and $5.0$. However, we have another case of a deceleration with $\tau_{0}=0.5$.}
	\label{rhosu}
  \end{center}
\end{figure}

The behavior of the deceleration parameter $q(\tau)$ can be clearly displayed in Fig.(\ref{rhosu}). Notice that the behavior of the deceleration parameter at very early time depends on $t_{0}$. Specifically, if we choose $t_{0}<2$ we find the positivity of the deceleration parameter at very early time. However, we clearly discover that the deceleration parameter is always negative at late time.

\begin{figure}[H]
\begin{center}
	\includegraphics[width=.6\textwidth]{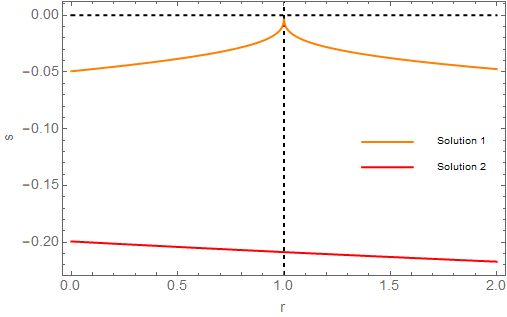}
	\caption{The plot shows the statefinder parameters in the $r-s$ plane. We notice that our model develops $(r, s)= (1, 0)$ for $\tau_{0}=2/3$ which displays the point where the flat $\Lambda$CDM model exists. Here we used $\rho_{0}=1$. }
	\label{rsrs}
  \end{center}
\end{figure}

Other physical quantities are the statefinder parameters $\left\{r,s\right\}$ defined in terms of the Hubble parameters as follow:
\begin{eqnarray}
r&=& 1 + 3\frac{\dot H}{H^2}+ \frac{\ddot{H}}{H^3}, \label{r02}\\
s&=& -\frac{3H\dot{H}+\ddot{H}}{3H\left( 2\dot{H}+3H^2\right)}. \label{s02}
\end{eqnarray}
In terms of $\tau$, they become
\begin{eqnarray}
r&=& 1 + \frac{3\partial_{\tau} H}{H^{5/2}}+ \frac{\partial_{\tau}(H^{-1/2}\partial_{\tau}H)}{H^{7/2}}, \label{r2}\\
s&=& - \frac{\partial_{\tau} H}{\left(2\partial_{\tau} H+3H^{5/2}\right)}- \frac{\partial_{\tau}(H^{-1/2}\partial_{\tau}H)}{3H\left(2\partial_{\tau} H+3H^{5/2}\right)}. \label{s2}
\end{eqnarray}
Substituting the Hubble parameter Eq.(\ref{H1}) into the above expressions, we find
\begin{eqnarray}
r&=&\frac{6 M(\tau)^{9/2} \rho_{0}^3\tau_{0}(3 \tau_{0}-2)+M(\tau)^3 \rho_{0}^6 (2-3 \tau_{0})^2+2 \tau_{0}^2}{2 \tau_{0}^2}, \label{r2h}\\
s&=& -\frac{(3 \tau_{0}-2) \left(6 N(\tau)^{3/2} \tau_{0}+3 \tau_{0}-2\right)}{6 N(\tau)^{3/2} \tau_{0}\left(3 \sqrt{N(\tau)} \tau_{0}+6 \tau_{0}-4\right)}, \label{s2h}
\end{eqnarray}
where we have used $\kappa=1$ and $H_{0}=1$ and defined new parameters $M(\tau)$ and $N(\tau)$ as follows:
\begin{eqnarray}
M(\tau)\equiv e^{3 \tau-\frac{2 \tau}{\tau_{0}}},\,\,N(\tau)\equiv M(\tau)/\rho_{0}^2. \label{r2h1}
\end{eqnarray}
We can write $\tau = \tau(s)$ by solving Eq.(\ref{s2h}) and substituting $t(r)$ back into Eq.(\ref{r2h}). Therefore we can obtain $r$ in terms of $s$, i.e., $r = r(s)$. Note that the statefinder parameters $\left\{r, s\right\} = \left\{1, 0\right\}$ represents  the point where the flat $\Lambda$CDM model exists in the $r-s$ plane \cite{huang}. Therefore the departure of dark energy models from this fixed point can be used to obtain  the distance of these models from the flat $\Lambda$CDM model. This allows us to display the statefinder parameters as shown in Fig.(\ref{rsrs}). Here we found two solutions in which the first one displays the existence of the $\Lambda$CDM model.
Notice that above physical quantities, e.g. the statefinder parameters, are parametrized by higher-order time derivatives of the scale factor. However, another diagnostic parameter depends only on the first order temporal  derivative of the scale factor can also be used to constrain the model. This is so-called the \emph{Om} analysis \cite{Sahni}. This parameter only involves the Hubble parameter. This has also been applied to some interesting models, e.g. Galileons models \cite{Jamil:2013yc}. The \emph{Om} analysis is defined in terms of the Hubble parameter via
\begin{eqnarray}
Om(z)=\frac{\left[\frac{H(z)}{H_0}\right]^2-1}{(1+z)^3-1},\label{Omz}
\end{eqnarray}
where \emph{Om}(z) is a function depending only on the redshift, $z$. Moreover, this can be parametrized further when having a constant equation of state (EoS) parameter $\omega$, and in this case we can write
\begin{eqnarray}
Om(z)= \Omega_{m0} + (1 -\Omega_{m0})\frac{(1+z)^{3(1+\omega)}-1}{(1+z)^3-1}.
\end{eqnarray}
It is worth noting that  we have different values of  $Om(z) =\Omega_{m0}$  for the  $\Lambda$CDM model, quintessence, and phantom cosmological models.
\begin{figure}[H]
\begin{center}
	\includegraphics[width=.47\textwidth]{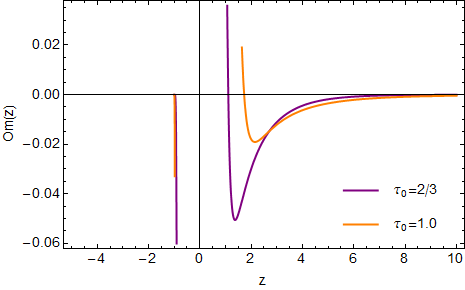}
    \includegraphics[width=.47\textwidth]{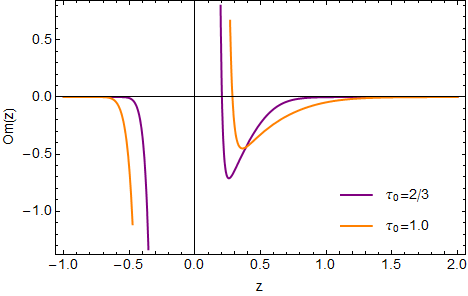}
	\caption{The plots show the relation between \emph{Om}$(z)$ and $z$ given by Eq.(\ref{Omz}). Here we have used $\rho_{0}=0.5$ and $\rho_{0}=1.0$ for the left panel and right panel, respectively.}
	\label{Omzz}
  \end{center}
\end{figure}
The redshift dependence of the \emph{Om}(z) parameter is displayed in Fig.(\ref{Omzz}). The plots show the relation between \emph{Om}$(z)$ and $z$ given by Eq.(\ref{Omz}). Here we have used $\rho_{0}=0.5$ and $\rho_{0}=1.0$ for the left panel and right panel, respectively.
\begin{figure}[H]
\begin{center}
	\includegraphics[width=.47\textwidth]{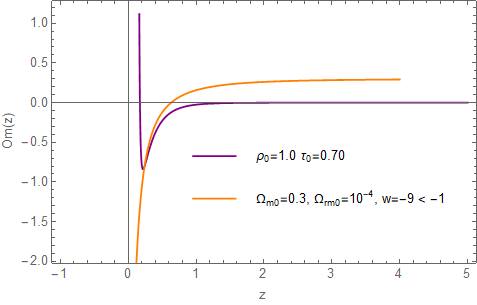}
    \includegraphics[width=.47\textwidth]{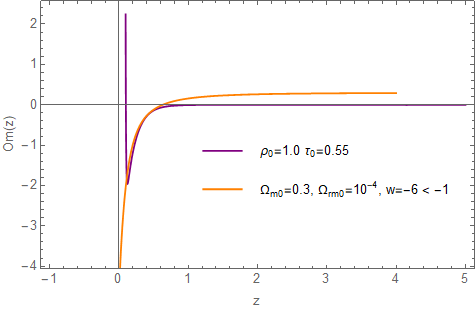}
	\caption{We compare the Hubble parameter of Eq.(\ref{42}) with that of Eq.(\ref{r02}). We find that they are in good agreement between $0.1 \leq z \leq 0.4$ for the left panel, while between $0.1 \leq z \leq 0.6$ for the right panel.}
	\label{4243}
  \end{center}
\end{figure}
From the left-panel of Fig.(\ref{4243}), we discover that our model displays an equation of state $w=-9$ which corresponds to the hypothetical phantom energy. Here we used for Eq.(\ref{42}) $\tau_{0}=1.0$ and $\rho_{0}=0.70$ and for Eq.(\ref{r02}) $\Omega_{m0}=0.3,\,\Omega_{rm0}=10^{-4}$ and $w=-9$. In addition, on the right-panel of Fig.(\ref{4243}), we consider an equation of state $w=-6$ which also corresponds to the hypothetical phantom energy. Here we used for Eq.(\ref{42}) $\tau_{0}=1.0$ and $\rho_{0}=0.55$ and for Eq.(\ref{r02}) $\Omega_{m0}=0.3,\,\Omega_{rm0}=10^{-4}$ and $w=-6$.

\section{Conclusion}
Nash gravity has proposed a new theory of gravity alternative to Einstein’s general theory of relativity. The formulation allows us to obtain field equations for empty space, but did not include a description of matter field. In the present work, we have generalized the original Nash theory by adding the matter fields in the original action. We have specified a proper form of the field equations on more general footings for space with matter contents. We have derived the equations of motion in the flat FLRW spacetime and examined the behaviors of the solutions by invoking specific forms of the Hubble parameter. 

We have also classified the physical behaviors of the solutions by employing the stability analysis. We have checked the consistency of the model by considering cosmological parameters, e.g., the Hubble parameter $H$, deceleration parameter $q$, and \emph{Om}(z) parameter. However, particular extensions of the present work are still possible. As well known, the dynamical systems analysis for analyzing the qualitative properties of cosmological models has proven to be very useful. It has been successfully used to study and to
understand a number of cosmological models, e.g. the standard GR cosmology \cite{sta} .

Moreover, there were some interesting topics left to be investigated regarding the less anisotropic counterpart (Bianchi types and the others); see Ref.\cite{Liu:2017edh} for example.  In summary we point out that  although this second order theory doesn't have classical Einstein limits, it has been proven to be formally divergent free and considered to be of interest in constructing theories of quantum gravity.



\section*{Acknowledgment}
P. Channuie is financially supported by the Institute for the Promotion of Teaching Science and Technology (IPST) under the project of the \lq\lq Research Fund for DPST Graduate with First Placement\rq\rq\,, under Grant No. 033/2557. The work of D. Momeni is supported by the Internal Grant (IG/SCI/PHYS/20/07) provided by Sultan Qaboos University.  The work of  M. Al Ajmi is supported by (IG/SCI/PHYS/20/02). 

\end{document}